\documentstyle[12pt]{article}
\begin{document}
\begin{center}

{\large\bf The topological AC effect on noncommutative phase space
} \vskip 1cm Kang Li$^{a, c,}$\footnote{kangli@hztc.edu.cn},
Jianhua Wang$^{b, c,}$\footnote{jianhuawang59@yahoo.com.cn}
\\\vskip 1cm

{\it\small$^a$ Department of Physics, Hangzhou Teachers College,
Hangzhou, 310036, P.R.China\\
$^b $Department of Physics, Shaanxi University  of Technology ,
Hanzhong, 723001,P.R. China}\\
$^c$The Abdus Salam International Center for Theoretical Physics,
Trieste, Italy
 \vskip 0.5cm
\end{center}

\begin{abstract}
The Aharonov-Casher (AC) effect in non-commutative(NC) quantum
mechanics  is studied. Instead of using the star product method,
we use a generalization of Bopp's shift  method. After solving the
Dirac equations both on noncommutative space and noncommutative
phase space by the new method, we obtain the corrections to AC
phase on NC space and  NC phase space respectively.

PACS number(s): 02.40.Gh,  11.10.Nx,  03.65.-w.
\end{abstract}

\section{Introduction}
In recent years, there has been an increasing interest in the
study of physical effects on non-commutative space, apart from
field theory, non-commutative quantum mechanics has recently
attract much attentions, because the effects of the space
non-commutativity may become significant in the extreme situation
such as in the string scale or at the Tev and higher energy level.
There are many papers devoted to the study of various aspects of
quantum mechanics on NC space with usual (commutative) time
coordinate \cite{1}-\cite{6}. For example, the topological AB  in
NC space and even NC phase space have studied in
\cite{2}-\cite{4}. In this work, other than the method employed in
\cite{5}, we develop a new method to obtain the corrections to the
topological phase of the AC effect both on NC space and NC phase
space, where we know that in a commutative space the line spectrum
does not depend on the relativistic nature of the dipoles. The
article is organized as follows: in section 2, by using the
Lagrangian formulation, we discuss the AC effect on a commutative
space in 2+1 dimensions. In section 3, the AC effect on a
noncommutative space is studied, and the correction to AC phase on
NC space is obtained,  the AC effect on a non-commutative phase
space will be discussed in section 4, and a most generalized
formula for holonomy in NC phase space is given in this section,
at last, some remarks are given in the last section.

\section{Description of AC effect in 2+1 commutative space time}

To begin with, let's  give a brief review  of AC effect in $2+1$
commutative space time.  The Lagrangian for a neutral particle of
spin-1/2 with an anomalous magnetic dipole moment $\mu_m$
interacting  with the electromagnetic field has the form
\begin{eqnarray}\label{eq6}
L = \bar \psi i \gamma^\mu \partial_\mu \psi - m\bar \psi \psi -
{1\over 2} \mu_m\bar \psi \sigma^{\mu\nu} \psi F_{\mu\nu}.
\label{ac}
\end{eqnarray}
The last term in the Lagrangian is responsible for the AC effect.

We restrict the particle moves on a plane (say $x-y$ plane), then
the problem can be treated in $2+1$ space time. We use the
following conventions for the 2+1 dimensional metric $g_{\mu\nu}$
and the anti-symmetric tensor $\epsilon_{\mu\nu\alpha}$:
\begin{equation}\label{eq7}
g_{\mu\nu} = \mbox{diag}(1,-1,-1) \quad \quad \mbox{and} \quad
\quad \epsilon_{012} = +1.
\end{equation}

Other than to use $2 \times 2$ matrices satisfying the 2+1
dimensional Dirac algebra, we will use $3$ four component Dirac
matrices which can describe spin up and down in the notional $z$
direction for a particle and for its anti-particle.  In 2+1
dimensions these Dirac matrices satisfy the following relation
\cite{7}:
\begin{equation}
\gamma^\mu \gamma^\nu = g^{\mu\nu}
-i\gamma^0\sigma^{12}\epsilon^{\mu\nu\lambda}\gamma_\lambda .
\end{equation}
 A particular representation is
\begin{eqnarray}
\gamma^0 &=& \left (\begin{array}{cc} \sigma_3&0\\
0&\sigma_3\end{array} \right ),~~ \gamma^1 = \left
(\begin{array}{cc} i\sigma_2&0\\0&-i\sigma_2
\end{array}\right ),~~ \gamma^2 = \left (\begin{array}{cc}
i\sigma_1&0\\ 0&i\sigma_1\end{array} \right ),
\end{eqnarray}

Then the interaction term in the Lagrangian can be written as
\begin{eqnarray}
 \bar \psi \sigma^{\mu\nu} \psi F_{\mu\nu} =
F^{\mu\nu}\gamma^0\sigma^{12}\epsilon_{\mu\nu\lambda} \bar \psi
\gamma^\lambda \psi,
\end{eqnarray}
with
\begin{eqnarray}
 F^{\mu\nu} = \left (
\begin{array}{ccc}
0&-E^1&-E^2\\ E^1&0&-B^3\\E^2&B^3&0 \end{array} \right ),
\label{dual}
\end{eqnarray}
where $E^i$ and $B^i$ are the electric and magnetic fields,
respectively. The indices ``1'' and ``2'' indicate the coordinates
on the $x-y$ plane along the $x$ and $y$ directions. The index
``3'' indicates the  $z$ direction.  The Lagrangian now can be
written as
\begin{eqnarray}
L =  \bar \psi i \gamma^\mu \partial_\mu \psi - m\bar \psi \psi
-(1/2)\gamma^0\sigma^{12}\mu_m \epsilon_{\mu\alpha\beta} F^{\alpha
\beta}\bar \psi \gamma^\mu \psi.
\end{eqnarray}

By using E-L equation, the Dirac equation of motion for a spin
half neutral particle with a magnetic dipole moment $\mu_m$  is
\begin{equation}\label{eq14}
(i\gamma_\mu
\partial^{\mu}-(1/2)\gamma^0\sigma^{12}\mu_m \epsilon_{\mu\alpha\beta} F^{\alpha
\beta}\gamma^\mu-m)\psi =0,
\end{equation}
and the solution will have the form
\begin{equation}\label{eq15}
\psi=e^{-\frac{i}{2}\gamma^0\sigma^{12}\mu_m \int^x
\epsilon_{\mu\alpha\beta} F^{\alpha \beta} dx^\mu}\psi_0,
\end{equation}
where $\psi_0$ is the solution for electromagnetic field free case
. The phase in Eq.(\ref{eq15}) is called AC phase , we write it as
\begin{equation}\label{eq16}
\phi_{AC}=-\frac{1}{2}\gamma^0\sigma^{12}\mu_m
\int^x\varepsilon_{\mu\alpha\beta}F^{\alpha\beta}dx^\mu .
\end{equation}
The  AC phase above is the general AB phase for a spin-1/2 neutral
particle passing through an electromagnetic field. If we consider
a situation of the standard AC configuration
\cite{AC}-\cite{Hagen}, i.e. the particle moves on a plane under
the influence of an pure electric field produced by a uniformly
charged infinitely long filament perpendicular to the plane,
 then we have
\begin{equation}\label{eq17}
\phi_{AC} = -\gamma^0\sigma^{12}\mu_m
\int^x\varepsilon_{0ij}F^{0i}dx^j = \gamma^0\sigma^{12}\mu_m
\int^x(\hat{\vec{k}}\times \vec{E})\cdot d\vec{x}=
\left(\begin{array}{cc}\sigma_3
&0\\0&-\sigma_3\end{array}\right)\int^x(\vec{\mu}_m\times
\vec{E})\cdot d\vec{x},
\end{equation}
where $\hat{\vec{k}}$ is the unit vector in $z$ direction and we
assume that the magnetic dipole moment is always along this
direction, i.e. $\vec{\mu}_m=\mu_m \hat{\vec{k}}$.

\section{ The AC Effect on noncommutative  space}

On the noncommutative space the coordinate and momentum operators
satisfy the following commutation relations (we take $\hbar=c=1$
unit)
\begin{equation}\label{eq1}
~[\hat{x}_{i},\hat{x}_{j}]=i\Theta_{ij},~~~
[\hat{p}_{i},\hat{p}_{j}]=0,~~~[\hat{x}_{i},\hat{p}_{j}]=i
\delta_{ij},
\end{equation}
where $\Theta_{ij}$ is an element of an antisymmetric matrix, it
is very small related to the energy scale and it represents the
non-commutativity of the NC space, $\hat{x}_i$ and $\hat{p}_i$ are
the coordinate and momentum operators on a NC space.

Just like the  static Schr$\ddot{o}$dinger equation on NC
space\cite{4}, the Dirac equation (\ref{eq14}) for a spin half
neutral particle with a magnetic dipole moment $\mu_m$, on NC
space, can be written as
\begin{equation}\label{Dirac}
(i\gamma_\mu
\partial^{\mu}-(1/2)\gamma^0\sigma^{12}\mu_m \epsilon_{\mu\alpha\beta} F^{\alpha
\beta}\gamma^\mu-m)\ast\psi =0,
\end{equation}
i.e, simply replace usual product with a star product (Moyal-Weyl
product), the Dirac equation in usual commuting space will change
into the Dirac equation on NC space. The star product between two
functions is defined by,
\begin{equation}\label{eq4}
(f  \ast g)(x) = e^{  \frac{i}{2}
 \Theta_{ij} \partial_{x_i} \partial_{x_j}
 }f(x_i)g(x_j)  = f(x)g(x)
 + \frac{i}{2}\Theta_{ij} \partial_i f \partial_j
 g\big|_{x_i=x_j}+{\mathcal{O}}(\Theta^2),
\end{equation}
here $f(x)$ and $g(x)$ are two arbitrary functions.

Some features of AC effect on noncommutative space has been
studied in \cite{5} by using the star calculation,  but it is
still meaningful to study it again by using the method gave in
reference \cite{4}, i.e. through a generalized Bopp$^\prime$s
shift, and the method can be easily generalized to NC phase space
which will be discussed in the next section.

On NC space the star product can be replaced by a Bopp's shift,
i.e the star product can be changed into ordinary product by
shifting coordinates $x_\mu$ with
\begin{equation}\label{eq19}
\hat{x}_\mu=x_\mu-\frac{1}{2}\Theta_{\mu\nu}p^\nu.
\end{equation}
Now, let us consider the noncommutative Dirac equation
(\ref{Dirac}), to replace the star product with ordinary product,
equivalent to the Bopp's shift, the $F_{\mu\nu}$ must, up to the
first order of the NC parameter $\Theta$, be shifted by
\begin{equation}\label{eq21}
F_{\mu\nu}\rightarrow\hat{F}_{\mu\nu}=F_{\mu\nu}+\frac{1}{2}
 \Theta^{\alpha\beta} p_\alpha\partial_\beta F_{\mu\nu}.
\end{equation}
Then  the Dirac equation for AC problem on NC space has the form
\begin{equation}\label{eq22}
(i\gamma_\mu
\partial^{\mu}-(1/2)\gamma^0\sigma^{12}\mu_m \epsilon_{\mu\alpha\beta} \hat{F}^{\alpha
\beta}\gamma^\mu-m)\psi =0.
\end{equation}
So, on NC space, the AC phase has the form:
\begin{eqnarray}\label{eq24}
\hat{\phi}_{AC}=-\frac{1}{2}\gamma^0\sigma^{12}\mu_m
\int^x\varepsilon_{\mu\alpha\beta}\hat{F}^{\alpha\beta}dx^\mu~~~~~~~~~~~~~~~~~~~~~~~~~~~~~~~~~~~~~~~~~~~~~~~~\nonumber\\
 =-\frac{1}{2}\gamma^0\sigma^{12}\mu_m
\int^x\varepsilon_{\mu\alpha\beta}F^{\alpha\beta}dx^\mu-\frac{1}{4}\gamma^0\sigma^{12}\mu_m\int^x\epsilon_{\mu\alpha\beta}\Theta^{\sigma\tau}
p_\sigma \partial_\tau F^{\alpha
  \beta}dx^\mu .
\end{eqnarray}
This is the general AC phase for a  spin-1/2 neutral particle
moving in a general electromagnetic field.

In the standard AC configuration, the momentum on NC space can be
written as \footnote{In the standard AC configuration the
Hamiltonian in commuting space has the form
\cite{AC}-\cite{Hagen}:
$$H=\frac{1}{2m}\vec{\sigma}\cdot(\vec{p}-i\mu_m
\vec{E})\vec{\sigma}\cdot(\vec{p}+i\mu_m \vec{E}).$$
In the region  $\vec{\nabla}\cdot \vec{E}=0$ (if the particle do
not reach at the charged filament, this is always  true), then the
equation above can be recast as
$$H=\frac{1}{2m}(\vec{p}- \vec{E}\times\vec{\mu})^2-\frac{\mu^2
E^2}{2m},$$
then the velocity operator can be gotten
$$v_l=\frac{\partial H}{\partial
p_l}=\frac{1}{m}[p_l-(\vec{E}\times\vec{\mu})_l].$$}
\begin{equation}\label{eq29}
p_l=mv_l+(\vec{E}\times\vec{\mu})_l+\mathcal{O}(\theta),
\end{equation}
where $\vec{\mu}=\mu_m\vec{\sigma}$, insert equation (\ref{eq29})
to (\ref{eq24}) and notice that
\begin{equation}\label{eq30}
F^{\alpha\beta}\longrightarrow F^{0i}~~ {\rm and}~~
 \Theta^{ij}=\theta\epsilon^{ij},~\Theta^{0\mu}=\Theta^{\mu 0}=0,
\end{equation}
we have
\begin{equation}\label{eq31}
\hat{\phi}_{AC}=\phi_{AC}+\delta\phi_{NCS},
\end{equation}
where $\phi_{AC}$ is the AC phase on commuting space given by
(\ref{eq17}), the  additional phase $\delta \phi_{NCS}$, related
to the non-commutativity of space, is given by
\begin{eqnarray}\label{eq32}
\delta\phi_{NCS}=-\frac{1}{2}\gamma^0\sigma^{12} \mu_m \int^x
\epsilon_{\mu0i}
\theta\epsilon^{\alpha\beta}[mv_\alpha+(\vec{E}\times
\vec{\mu})_\alpha]\partial_\beta F^{0i}dx^\mu ~~~~~~~~\nonumber\\
=\frac{1}{2}\gamma^0\sigma^{12}
\mu_m\theta\epsilon^{ij}\int^x[k_j+(\vec{E}\times \vec{\mu})_j]
(\partial_iE^2dx^1-\partial_iE^1dx^2),~~
\end{eqnarray}
where $k_j=mv_j$, and the result here coincides with the result
  given in reference \cite{5}, where the tedious star product calculation has been used.

  The first term is a velocity dependent correction and
does not have the topological properties of the commutative AC
effect and could modify the phase shift. The second term is a
correction to the vortex and does not contribute to the line
spectrum.

\section{The AC effect on noncommutative phase space}

We have discussed the AC effect on  NC space, where space-space do
not commute but with momentum-momentum commuting. The
Bose-Einstein statistics in noncommutative quantum mechanics
requires both space-space and momentum-momentum non-commutativity,
we call the NC space with momenta non-commuting is NC phase space.
So study the physics on NC phase space is very important. On NC
phase space, the momentum $\hat{p_i}$'s commutation relation in
(\ref{eq1}) should be replaced with
\begin{equation}
[\hat{p}_{i},\hat{p}_{j}]=i\bar{\Theta}_{ij},
\end{equation}
where $\bar{\Theta}$ is the antisymmetric matrix, its elements
represent the non-commutativity of the momenta. The Dirac equation
(\ref{eq14}) on NC phase space can also be written as:
\begin{equation}\label{DiracEqu3}
(-\gamma_\mu p^{\mu}-(1/2)\gamma^0\sigma^{12}\mu_m
\epsilon_{\mu\alpha\beta}F^{\alpha \beta}\gamma^\mu-m)\ast\psi =0,
\end{equation}
 but here,
the star product in Eqs. (\ref{DiracEqu3}) defines,
\begin{eqnarray}
(f  \ast g)(x,p) = e^{ \frac{i}{2\alpha^2}
 \Theta_{ij} \partial_i^x \partial_j^x+\frac{i}{2\alpha^2}\bar{\Theta}_{ij} \partial_i^p
 \partial_j^p}
 f(x,p)g(x,p) ~~~~~~~~~~~~~~~~~~~~~~~\nonumber\\ = f(x,p)g(x,p)
 + \frac{i}{2\alpha^2}\Theta_{ij} \partial_i^x f \partial_j^x g\big|_{x_i=x_j}
 + \frac{i}{2\alpha^2}\bar{\Theta}_{ij} \partial_i^p f \partial_j^p
 g\big|_{p_i=p_j}+{\mathcal{O}}(\Theta^2),
\end{eqnarray}
where ${\mathcal{O}}(\Theta^2)$ stands the second and higher order
terms of $\Theta$ and $\bar{\Theta}$. The star product in Dirac
equation on NC phase space can be placed by the usual product from
the two steps, first we need to replace $x_i$ and $ p_i$   with a
generalized Bopp's shift as
\begin{eqnarray}\label{gbshift1}
x_\mu\rightarrow \alpha x_{i}-\frac{1}{2 \alpha}\Theta_{\mu\nu}p_{\nu},\nonumber\\
p_\mu\rightarrow \alpha
p_\mu+\frac{1}{2\alpha}\bar{\Theta}_{\mu\nu}x_{\nu},
\end{eqnarray}
and also need the partner of shift in  Eq.(\ref{eq21}) in NC phase
space as,
\begin{equation}\label{LW-shift}
F_{\mu\nu}\rightarrow\hat{F}_{\mu\nu}=\alpha
F_{\mu\nu}+\frac{1}{2\alpha}
 \Theta^{\alpha\beta} p_\alpha\partial_\beta F_{\mu\nu}.
\end{equation}
The Dirac equation (\ref{DiracEqu3}) then read
\begin{equation}\label{DiracEqu4}
\{-\alpha\gamma^\mu
p_{\mu}-\frac{1}{2\alpha}\gamma^\mu\bar{\Theta}_{\mu\nu}x_{\nu}-(1/2)\gamma^0\sigma^{12}\mu_m
\epsilon_{\mu\alpha\beta}[\alpha F^{\alpha\beta}+\frac{1}{2\alpha}
 \Theta^{\tau\sigma} p_\tau\partial_\sigma F^{\alpha\beta}]\gamma^\mu-m \}\psi =0.
\end{equation}
Because $\alpha\neq 0$, so the above Dirac equation can be recast
to
\begin{equation}\label{DiracEqu5}
\{-\gamma^\mu
p_{\mu}-\frac{1}{2\alpha^2}\gamma^\mu\bar{\Theta}_{\mu\nu}x_{\nu}-(1/2)\gamma^0\sigma^{12}\mu_m
\epsilon_{\mu\alpha\beta}[F^{\alpha\beta}+\frac{1}{2\alpha^2}
 \Theta^{\tau\sigma} p_\tau\partial_\sigma F^{\alpha\beta}]\gamma^\mu-m' \}\psi
 =0,
\end{equation}
where $m'=m/\alpha$. The solution to (\ref{DiracEqu5}) is
\begin{equation}
\psi=e^{i\hat{\varphi}_{AC}}\psi_0,
\end{equation}
where $\psi_0$ is the solution of Dirac equation for free particle
with mass $m'$, and the $\hat{\varphi}_{AC}$ stands AC phase in NC
phase space, and it has the form below,
\begin{eqnarray}\label{phase-NCPhase-space}
\hat{\varphi}_{AC}=-\frac{1}{2}\gamma^0\sigma^{12}\mu_m
\int^x\varepsilon_{\mu\alpha\beta}F^{\alpha\beta}dx^\mu~~~~~~~~~~~~~~~~~~~~~~~~~~~~~
\nonumber\\-\frac{1}{2\alpha^2 }\int^x \bar{\Theta}_{ij}x_jdx_i-
\frac{1}{4\alpha^2}\gamma^0\sigma^{12}\mu_m\int^x\epsilon_{\mu\alpha\beta}\Theta^{\sigma\tau}
p_\sigma \partial_\tau F^{\alpha
  \beta}dx^\mu.
\end{eqnarray}
Equation (\ref{phase-NCPhase-space}) is the general AC phase in
noncommutative phase space. For the standard AC case i.e. particle
moves in a pure  static electric field, then the AC phase reduces
to
\begin{eqnarray}\label{phase-NCPhase-space2}
\hat{\varphi}_{AC}=\phi_{AC}-\frac{1}{2\alpha^2 }\int^x
\bar{\Theta}_{ij}x_jdx_i-\frac{1}{2\alpha^2}\gamma^0\sigma^{12}
\mu_m \int^x \epsilon_{\mu0i}
\theta\epsilon^{\alpha\beta}[m'v_\alpha+(\vec{E}\times
\vec{\mu})_\alpha]\partial_\beta F^{0i}dx^\mu ~~~~~~~~\nonumber\\
=\phi_{AC}-\frac{1}{2\alpha^2 }\int^x
\bar{\Theta}_{ij}x_jdx_i+\frac{1}{2\alpha^2}\gamma^0\sigma^{12}
\mu_m\theta\epsilon^{ij}\int^x[k'_j+ (\vec{E}\times \vec{\mu})_j]
(\partial_iE^2dx^1-\partial_iE^1dx^2),~~
\end{eqnarray}
where $k'_j=m'v_j, ~p_l=m'v_l+
(\vec{E}\times\vec{\mu})_l+\mathcal{O}(\theta)$ has been applied
and we omit the second order terms in $\theta$. Equation
(\ref{phase-NCPhase-space2}) can also be written as
\begin{equation}
\hat{\varphi}_{AC}=\phi_{AC}+\delta\phi_{NCS}+\delta\phi_{NCPS},
\end{equation}
where $\phi_{AC}$  is the AC phase on commuting space ( see Eq.
\ref{eq17}), $\delta\phi_{NCS}$ is the space-space non-commuting
contribution to the AC phase on NC space (see Eq. \ref{eq32}), and
the last term $\delta\phi_{NCPS}$ is given by
\begin{eqnarray}\label{phase-NCPhase-space3}
\delta\phi_{NCPS}=-\frac{1}{2\alpha^2 }\int^x
\bar{\Theta}_{ij}x_jdx_i+\frac{1-\alpha^3}{2\alpha^3}\gamma^0\sigma^{12}
\mu_m\theta\epsilon^{ij}\int^x
k_j(\partial_iE^2dx^1-\partial_iE^1dx^2)\nonumber\\+
\frac{1-\alpha^2}{2\alpha^2}\gamma^0\sigma^{12}
\mu_m\theta\epsilon^{ij}\int^x(\vec{E}\times \vec{\mu})_j
(\partial_iE^2dx^1-\partial_iE^1dx^2),
\end{eqnarray}
which represents the non-commutativity of the momenta. When
$\alpha=1$, which will lead to $\bar{\Theta}_{ij}=0$ \cite{6},
then the AC phase on NC phase space will return to the AC phase on
NC space, i.e. $\delta\phi_{NCPS}=0$ and equations
(\ref{phase-NCPhase-space}) and (\ref{phase-NCPhase-space2}) will
change into equations (\ref{eq24}) and (\ref{eq31}) respectively.

\section{Conclusion remarks }

In this paper, we study the AC effect on both noncommutative space
and noncommutative phase space. Instead of doing tedious star
product calculation, we use the "shift" method, i.e. the star
product in Dirac equation can be replaced by Bopp's shift and
together with the shift we defined in (\ref{eq21}) for NC space
and (\ref{LW-shift}) for NC phase space. These shifts is exact
equivalent to the star product. The additional AC phase
(\ref{eq32}) in NC space is exact the same as in reference
\cite{5} shows the correctness of our method. Our results of AC
phase in NC phase space, especially the new term
(\ref{phase-NCPhase-space3}) which comes from the momenta
non-commutativity is totally new result of this paper.

The method we use in this paper may also be employed to other
physics problem on NC space and NC phase space. The further study
on the issue will be reported in our forthcoming papers.

\section{Acknowledgments} This paper was completed during our visit
to the high energy section of the Abdus Salam International Centre
for Theoretical Physics (ICTP). We would like to thank Prof. S.
Randjbar-Daemi for his kind invitation and warm hospitality during
our visit at the ICTP. This work is supported in part by the
National Natural Science Foundation of China ( 10575026, and
10447005). The authors also grateful to the support from the Abdus
Salam ICTP, Trieste, Italy.

\end{document}